\documentclass[twocolumn]{aastex63}
\usepackage{textcomp}

\def\lea{\mathrel{<\kern-1.0em\lower0.9ex\hbox{$\sim$}}}
\def\gea{\mathrel{>\kern-1.0em\lower0.9ex\hbox{$\sim$}}}

\received{19 February, 2021}
\revised{30 May, 2021}
\accepted{10 June, 2021}
\submitjournal{\textsc{The Astrophysical Journal}}

\shorttitle{The Star Formation History of a Post-Starburst Galaxy}
\shortauthors{Chandar et al.}

\begin{document}

\title{The Star Formation History of a Post-Starburst Galaxy Determined From Its Cluster Population}
\author{Rupali Chandar}
\affil{Ritter Astrophysical Research Center, The University of Toledo, Toledo, OH 43606, USA}
\author[0000-0001-7413-7534]{Angus Mok}
\affil{Ritter Astrophysical Research Center, The University of Toledo, Toledo, OH 43606, USA}
\author[0000-0002-4235-7337]{K. Decker French}
\affil{Department of Astronomy, University of Illinois, Urbana IL, 61801, USA}
\author[0000-0003-2599-7524]{Adam Smercina}
\affil{Astronomy Department, University of Washington, Seattle, WA 98195, USA}
\author[0000-0003-1545-5078]{John-David T. Smith}
\affil{Ritter Astrophysical Research Center, The University of Toledo, Toledo, OH 43606, USA}

\correspondingauthor{Rupali Chandar}
\email{Rupali.Chandar@utoledo.edu}
\begin{abstract}
    The recent star formation histories (SFH) of post-starburst galaxies have been determined almost exclusively from detailed modeling of their composite star light.  This has provide important but limited information on the number, strength, and duration of bursts of star formation.  In this work, we present a direct and independent measure of the recent SFH of post-starburst galaxy S12 (plate-mjd-fiber for SDSS 623-52051-207; designated EAS12 in \citealt{Smercina2018}) from its star cluster population.  We detect clusters from high resolution, $UBR$ optical images taken with the {\em Hubble Space Telescope}, and compare their luminosities and colors with stellar population models to estimate the ages and masses of the clusters.  No clusters younger than $\sim70$~Myr are found, indicating star formation shut off at this time.  Clusters formed $\sim120$~Myr ago reach masses up to $\sim \mbox{few}\times10^7~M_{\odot}$, several times higher than similar age counterparts formed in actively merging galaxies like the Antennae and NGC~3256.  We develop a new calibration based on known properties for 8 nearby galaxies to estimate the star formation rate (SFR) of a galaxy from the mass of the most massive cluster, $M_{\rm max}$.  The cluster population indicates that S12 experienced an extremely intense but short-lived burst $\sim120$~Myr ago, with an estimated peak of $500^{+500}_{-250}~M_{\odot}~\mbox{yr}^{-1}$ and duration of $50\pm25$~Myr, one of the highest SFRs estimated for any galaxy in the modern universe. The cluster population also allows us to fill in more of the backstory of S12. Prior to the recent, intense burst, S12 was forming stars at a moderate rate of $\sim 3-5~M_{\odot}~\mbox{yr}^{-1}$, typical of spiral galaxies, but the system experienced an earlier burst at some point, approximate $1-3$~Gyr ago.  While fairly uncertain, we estimate that the SFR during this earlier burst was $\sim20-30~M_{\odot}~\mbox{yr}^{-1}$, similar to the current SFR in the Antennae and NGC~3256. \\
\end{abstract}

\section{Introduction} \label{sec:intro}
Contemporary models of galaxy evolution require feedback or other processes to expel most of the gas and dust fueling stellar and black hole growth, driving galaxies into quiescence \citep[e.g.,][]{Hopkins14,Correa19,Terrazas20}.  Post-starburst galaxies (PSBs) are systems found exactly during this transitional phase, shortly after star formation has been abruptly halted. Although rare today, estimates suggest that up to 40\% of all quiescent galaxies were rapidly quenched at some point in their lives \citep{Snyder2011,Wild2016}. In fact, star formation in these galaxies is believed to have shut off so rapidly that they must have been subjected to some of the most extreme feedback experienced by galaxies. Despite this, much remains unknown about the physical processes which create PSBs, and current cosmological simulations have a hard time producing PSBs amongst reasonable `quenched' galaxy populations \citep[e.g.][]{schaye2015,spilker2018}. The detailed properties of PSBs' past star-forming bursts, including timing, intensity, and duration, provide important information for fine-tuning the prescriptions used for feedback in galaxy simulations.

The disturbed morphologies, shells and tidal features obvious in many post-starbursts are direct evidence that these systems experienced a recent major merger \citep[e.g.][]{Zabludoff96, Yang04, Yang08, Pawlik15}.  Simulations of merging galaxies predict that one or two exponentially declining starburst events will be triggered during the interaction.  These simulations have been used to guide fits of the recent star formation histories (SFHs) of post-starburst galaxies. \citet{French18} fit composite star-light observations (far- and near- ultraviolet photometry from GALEX, broad-band SDSS $ugriz$ optical photometry and Lick indices measured from optical, SDSS spectra), which assume: (1) an old stellar population represented by a  linear-exponential  starting 10~Gyr ago, and (2) one or two exponential declining bursts which took place in the last Gyr.  

A direct and independent method of measuring the recent SFH (last $\sim0.5$~Gyr) in galaxies is to use the ages and masses of their bright stellar clusters — a method which so far has been relatively unexplored in post-burst galaxies, but has been successfully applied to the cluster populations in a broad range of nearby star-forming systems, including ongoing mergers. Because they are bright, single stellar populations, clusters can be studied even in fairly distant galaxies with $HST$, which is the only facility capable of detecting these populations at the distances of post-starbursts.  Young stellar clusters directly trace the star formation process; an increase in the rate of star formation results in an increase in the total number and mass of the most massive clusters \citep[e.g.][]{Chandar17, Whitmore20}. 

There have been very few studies of the cluster populations in post-starbursts to date. The most comprehensive, by \citet{Yang08}, used optical imaging in two filters ($B$ and $R$) with $HST$, and detected cluster populations in 5 PSBs. They found that the brightest of these clusters are brighter than almost all ancient globular clusters in the Milky Way, as expected for clusters formed within the last Gyr in merging systems. They were also able to provide broad constraints on the cluster ages and hence on the starburst event, from the $B-R$ colors of 5 to 10 clusters in each system, which indicated these events occurred more than 10~Myr ago, but more recently than a Gyr.

\begin{figure*}[!ht]
	\centering
	\includegraphics[width=6.in]{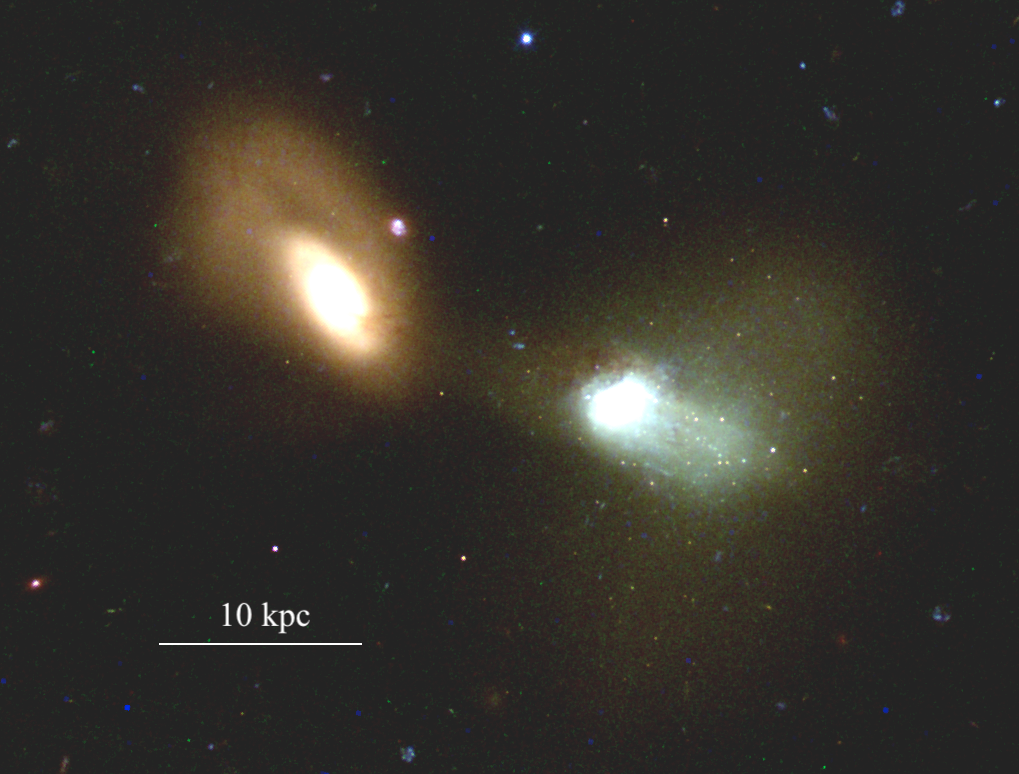}
	\caption{Post-starburst galaxy S12 and its companion to the northeast, in an \textit{HST UBR} composite image (asinh-scaled). The difference in starlight color reflects the different age stellar populations which dominate each galaxy, with S12 appearing more `green' due to recently formed stars and clusters, and its companion appearing yellow-orange because it contains an ancient stellar population. }\label{fig:s12fov}
\end{figure*}

In this work, we use archival $HST$ observations of the post-starburst system S12 (SDSS 623-52051-207), and its more massive companion (SDSS 6316-56483-722), shown in Figure~\ref{fig:s12fov}, to detect and study its star cluster population. This system is composed of two interacting galaxies, one to the Northeast which shows tidal distortions but is dominated by old stellar populations, and a significantly younger galaxy to the southwest which shows a 'jelly-fish' like extension and extended tidal features, which we refer to as S12. This system has one of the youngest fit ages in the post-starburst sample studied by \citet{French18}, who find a single burst which ended $\sim77$~Myr ago and had a duration of 25~Myr based on fits to the composite starlight. The main goal of this work is to independently determine the star formation history of S12 from its star cluster population.

The rest of this paper is organized as follows. Section~\ref{sec:obs} summarizes the $HST$ observations, cluster detection and photometry. Section~\ref{sec:analysis} presents the luminosity and color distributions of the clusters, and compares them with single stellar population model predictions to estimate the age and mass of each cluster. In Section~\ref{sec:Results} we develop a new method to estimate SFR from the most massive cluster, based on a calibration of well-known, nearby cluster populations in galaxies with a range of SFRs; we use this new calibration to estimate the star formation rate of S12 in different time intervals. In Section~\ref{sec:discussion} we present the star formation history of S12 and compare it with that of other types of galaxies, including on-going merging systems. We also compare our results with those from composite starlight fitting. Finally, in Section~\ref{sec:conclusion} we summarize our main results.

\section{Observations, Detection, \& Photometry} \label{sec:obs}
Post-starburst galaxy S12 is interacting with a nearby companion (see Figure~\ref{fig:s12fov}), and has a `jellyfish'-like structure, with stellar emission pointing to the southwest (away from the companion). It has an absorption line spectrum dominated by A-stars, and no detected line emission. S12 has a measured redshift of $z=0.0336$, or a luminosity distance of 148~Mpc; we adopt a distance modulus $m-M= 35.85$ here. The estimated stellar mass of S12 is somewhat uncertain because this system has an unusual star formation history, with stars formed at two distinct ages (rather than continuously) dominating the optical and near-infrared emission (see Section~\ref{subsec:formation}). From near-infrared measurements of the total luminosity in the WISE 1 band and Spitzer 3.6$\mu$m imaging, we estimate a stellar mass between $\sim1-3\times10^{9}~M_{\odot}$ where we have assumed a range of M$/$L ratios appropriate for the young/intermediate age stellar populations. From SDSS r-band photometry, the range is $\sim0.9 - 7\times 10^{9}~M_{\odot}$, based on assuming a young (100~Myr) or intermediate age ($\sim2\pm1$~Gyr) stellar population. The companion to the northeast appears to only have an ancient stellar population, including a few ancient clusters. We compare the properties of clusters around the companion galaxy with those of S12 when appropriate, since these provide a good control sample of what an old globular cluster population looks like at this distance.

S12 is the youngest post-starburst observed as part of a larger $HST$ survey (program GO-11643; PI: Zabludoff), which targeted systems covering a wide range of post-starburst ages (estimated from fitting their composite light) and also had pre-existing \textit{Spitzer} observations. It is one of the few post-starbursts imaged in three (rather than two) optical filters: F336W ($U$), F438W ($B$), F625W ($R$) filters, sufficient to estimate the ages of the clusters. We downloaded the reduced WFC3/UVIS images in the $U$, $B$, and $R$ bands from the Hubble Legacy Archive\footnote{http:$//$hla.stsci.edu}, which corrects each exposure for dark current and bias, then flat-fields and drizzles them together through the standard WFC3 pipeline to a scale of $0.04\arcsec$~pix$^{-1}$. Basic information on the observations are summarized in Table~\ref{tab:obs}. We present a three-color optical image of S12 in Figure~\ref{fig:s12fov}, which clearly shows the bright central nuclear region, extended diffuse stellar light, and many individual stellar clusters.

At the distance of S12, stellar clusters are unresolved point sources. Because the point-like clusters are embedded in diffuse stellar emission, we run a detection algorithm (\texttt{Photutils}\footnote{Based on the DAOFIND algorithm \citep{daophot}}; \citealt{photutils}) on a `median divided image', where the original $V$ band image is divided by a smoothed version to `flatten' the strongly varying background emission. S12 is too far away to detect individual stars older than $\sim10$~Myr, so potential contamination in our source list comes primarily from background galaxies. We eliminate 2 background galaxies based on visual inspection; our final catalog contains 115 candidate clusters.

\begin{figure*}[t]
	\centering
	\includegraphics[width=6.in]{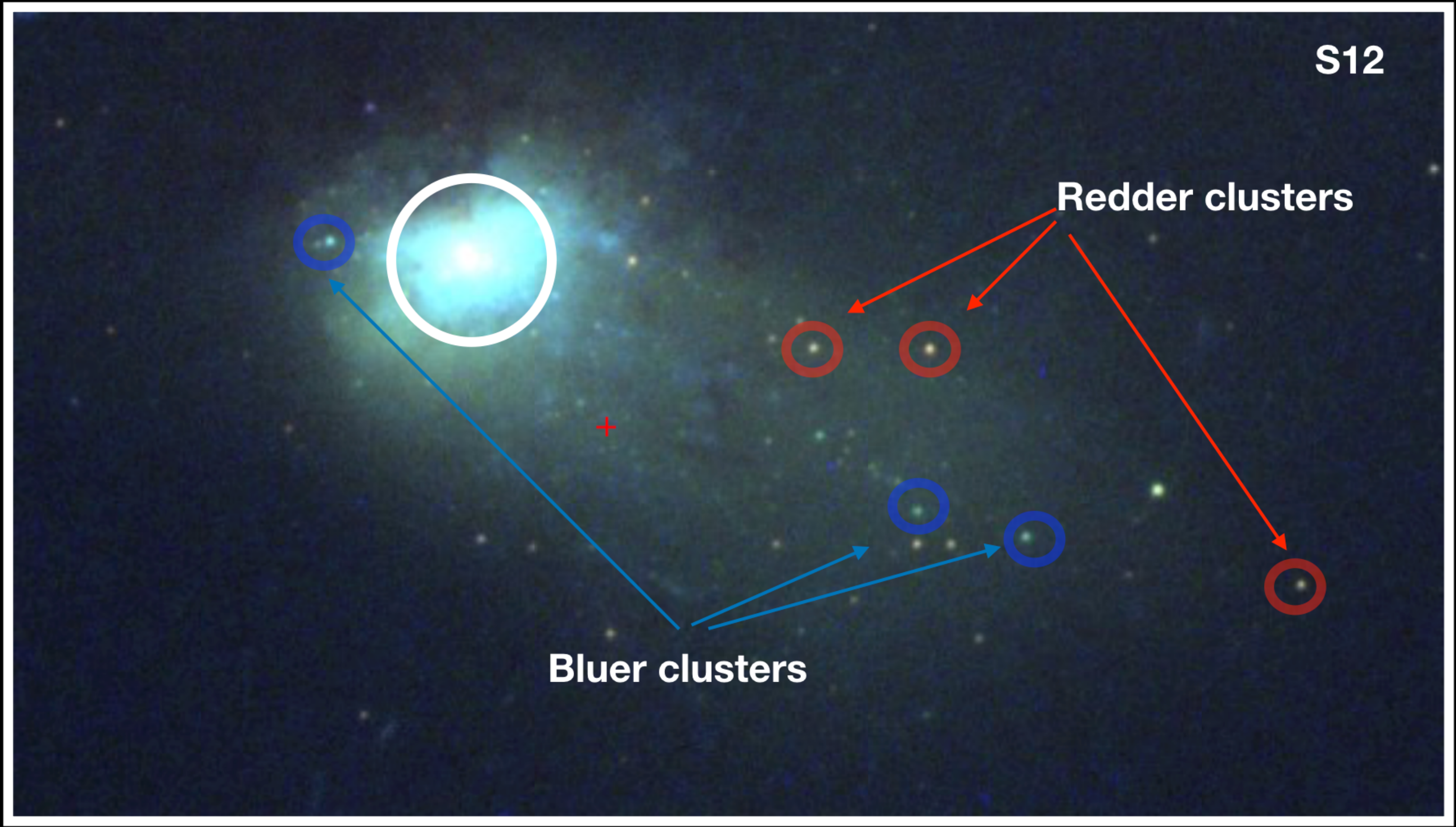}
	\caption{A zoomed-in, $UBR$ color image of S12 from $HST$ observations. A number of point-like clusters can clearly be seen in the image; we identify a few that obviously have redder and bluer colors; The differences in color are due to differences in age rather than differential reddening by dust. The white circle represents the $3\arcsec$ SDSS fiber used to obtain composite light optical spectra of the system. }\label{fig:S12}
\end{figure*}

We measure the brightness of each cluster in a 2-pixel aperture, using an annulus between 7 and 9 pixels to estimate the local background and subtract it off. We find that this small aperture gives slightly less scatter in the measured colors of the clusters relative to the model predictions (see next section), since it minimizes contamination from the diffuse starlight surrounding many of the clusters. We apply an aperture correction of $-0.65$~mag to each filter, indicated by the enclosed energy percentage in a 2-pixel radius published on the WFC3 instrument page. We measure the magnitudes of a several of the brighter, more isolated sources in 2 and 10 pixel radii in the R-band image, confirm this value, and estimate that the uncertainties are on the order of $\pm0.1$. Finally, we convert the instrumental magnitudes to the VEGAMAG system by applying the following zeropoints from the WFC3 data handbook to our measured photometry: 23.46 ($U$), 24.98 ($B$), and 25.37 ($R$)~mag. We do not make any correction to our measured magnitudes for foreground extinction, since it is quite low towards S12 ($A_V=0.064$); this assumption has very little impact on our age-dating results (see Section~\ref{subsec:massage}).

The clusters in our sample have a range of apparent R band magnitudes from $\sim20.3-27.3$~mag, which corresponds to absolute R band magnitudes of $\sim-15.5$ to $-8.5$~mag (assuming a distance modulus $m-M=35.85$~mag). As we will see in Section~\ref{sec:Results}, this means we can detect clusters with masses down to $\sim10^4~M_{\odot}$ at ages younger than 10~Myr. We also find that the clusters in S12 have a range of colors, and have identified a few of the redder and bluer ones in Figure~\ref{fig:S12}. As explained in Sections~3.2 and 3.3, these colors reflect differences in the ages of the clusters rather than differences in reddening. We estimate the photometric uncertainty for each source as follows. $\sigma_{\rm err} = \sqrt{\sigma^2_{\rm bkg} + \frac{I}{g_{\rm eff}}}$ where $\sigma_{\rm bkg}$ is the background counts, $I=$total source counts, $g$ is the effective gain. The formal errors are smallest in the $R$ band, and largest in $U$, with ranges from $\sim0.01 - 0.05$~mag for the former, and $\sim0.02-0.3$ for the latter. The brightest, blue clusters have higher uncertainties than expected just from source counts because the reside in regions where the background is strong and changes quickly, but are still less than $\sim0.05$~mag. We present our final cluster catalog, including positions and $UBR$ photometry, in Table~\ref{tab:clustercat}.

\begin{table}[t]
	\caption{Summary of HST/UVIS Observations of Post-starburst Galaxy S12}\label{tab:obs}
	\begin{tabular}{lcc}
	    \hline\hline
		Filter & \# Exposures & Total Exp  \\
		    &  & Time (s) \\ 
		\hline
		F336W ($U$) & $6\times998$s & 5,988   \\
		F438W ($B$) & $6\times855$s & 5,130  \\
		F625W ($R$) & $6\times454$s & 2,724  \\
		\hline
	\end{tabular}
\end{table}

\begin{table*}[ht]
	\caption{Catalog of Star Clusters in Poststarburst Galaxy S12}\label{tab:clustercat}
	\centering
	\begin{tabular}{lccccc}
	    \hline\hline
		Source & RA  & DEC & U band & B band & R band  \\
		  ID  & (J2000)  & (J2000) & (mag) & (mag) & (mag) \\ 
		\hline
		1 & 243.375330 & 51.059272 & $27.47\pm0.05$ & $27.62\pm0.03$ &  $26.85\pm0.02$  \\
		2 & 243.375104 & 51.059331 & $26.44\pm0.04$ & $26.81\pm0.02$ &  $26.39\pm0.02$  \\
		3 & 243.376134 & 51.059418 & $26.88\pm0.04$ & $26.95\pm0.02$ &  $26.55\pm0.02$  \\		
		\hline
	\end{tabular}
\end{table*}

\section{Analysis} \label{sec:analysis}
Two of the most basic properties of a cluster are its age and mass. These properties can be determined by comparing the measured colors and luminosities of the clusters with predictions from stellar population models. In this Section, we present the color-magnitude and color-color distributions of clusters in S12, and then estimate their individual ages and masses.  We also present cluster luminosity functions.

\subsection{Color-Magnitude Diagram and Luminosity Function} \label{subsec:cmd}
\begin{figure*}[t]
	\centering
	\includegraphics[width=5.5in]{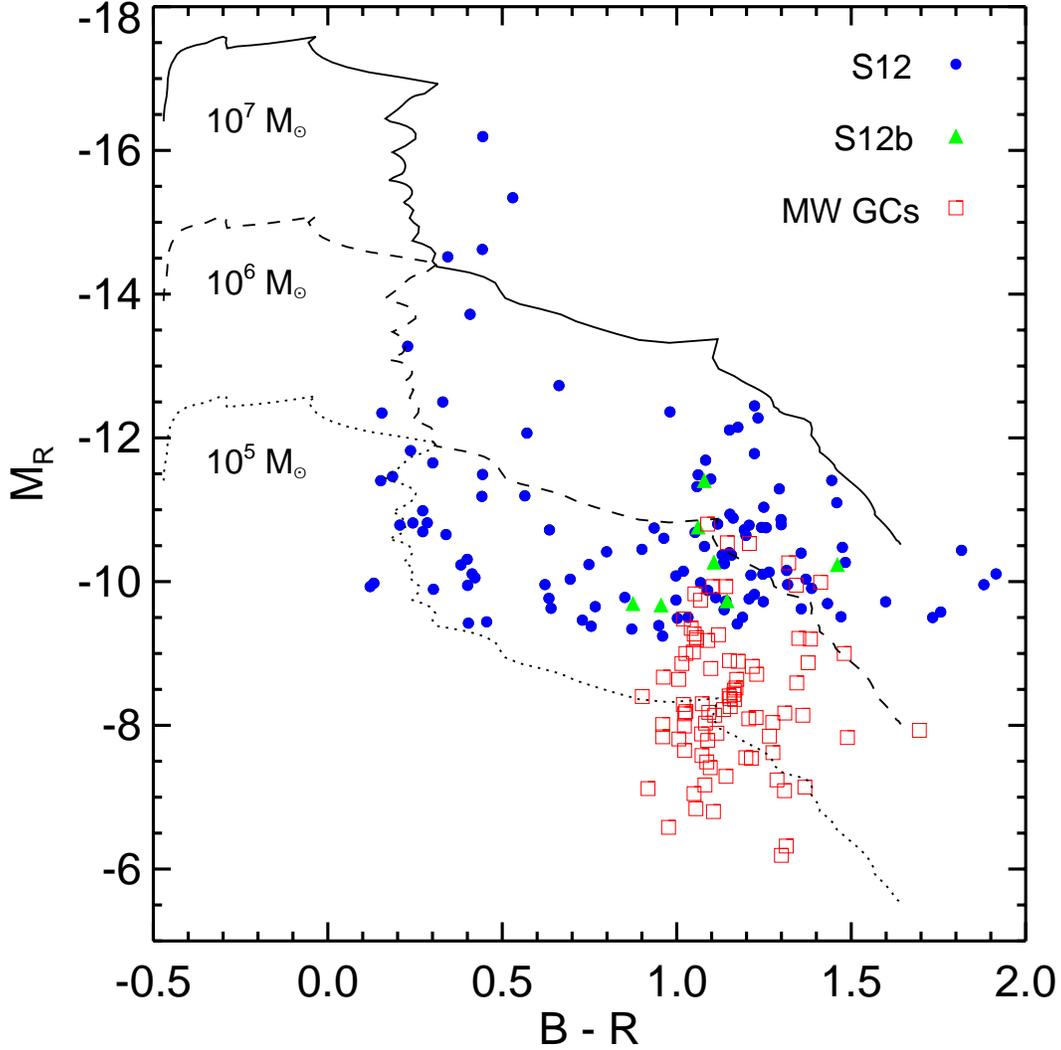}
	\caption{Clusters detected in S12 (blue, filled circles) show a range of magnitudes and colors in a $M_R$ vs. $B-R$ color-magnitude diagram. The seven clusters detected in the companion (green, filled triangles) and Galactic globular clusters (red, open squares) have redder colors and fainter magnitudes than their counterparts in S12. The lines show the predicted fading over time from top to bottom for a $10^7$ (solid curve), $10^6$ ( dashed curve), and a $10^5~M_{\odot}$ cluster (dotted curve).}\label{fig:cmd}
\end{figure*}

Clusters fade and become redder over time, as the most massive stars die off. This is illustrated by the curves in Figure~\ref{fig:cmd}, which track changes in $R$-band luminosity and $B-R$ color over the full lifespan of clusters, starting at $1$~Myr and extending to $\sim13$~Gyr. Each curve represents the predicted evolution for a cluster of a given mass: $10^7$, $10^6$, and $10^5~M_{\odot}$, from top left to bottom right. We can see that there is a population of extremely massive blue clusters in S12 (represented by the filled circles), with a few exceeding $10^7~M_{\odot}$. Some of the redder clusters with $B-R$ colors between 1.0 and 1.5 also appear to be quite massive (with $M>10^6~M_{\odot}$), although not quite as massive as their blue counterparts.

The handful of clusters detected in S12b, the interacting companion galaxy (shown as the open circles), all have similar colors to those of the reddest clusters in S12, but significantly fainter luminosities. For comparison, we also show the dereddened $B-R$ colors and $R$-band luminosities of ancient globular clusters in the Milky Way (red squares). The colors of these ancient stellar systems overlap with the red clusters in S12 and all clusters detected in S12. However, the brightest Galactic globular clusters are several orders of magnitude fainter than the red clusters in the post-starburst galaxy S12 (but quite similar to the brightest clusters in the companion S12b). This suggests that the clusters in the companion galaxy S12b are a typical, old globular cluster system, but that the red clusters in S12 are not; we will return to this point in Section~\ref{sec:oldsfh}.

\begin{figure*}[t]
	\centering
	\includegraphics[width=12.5cm]{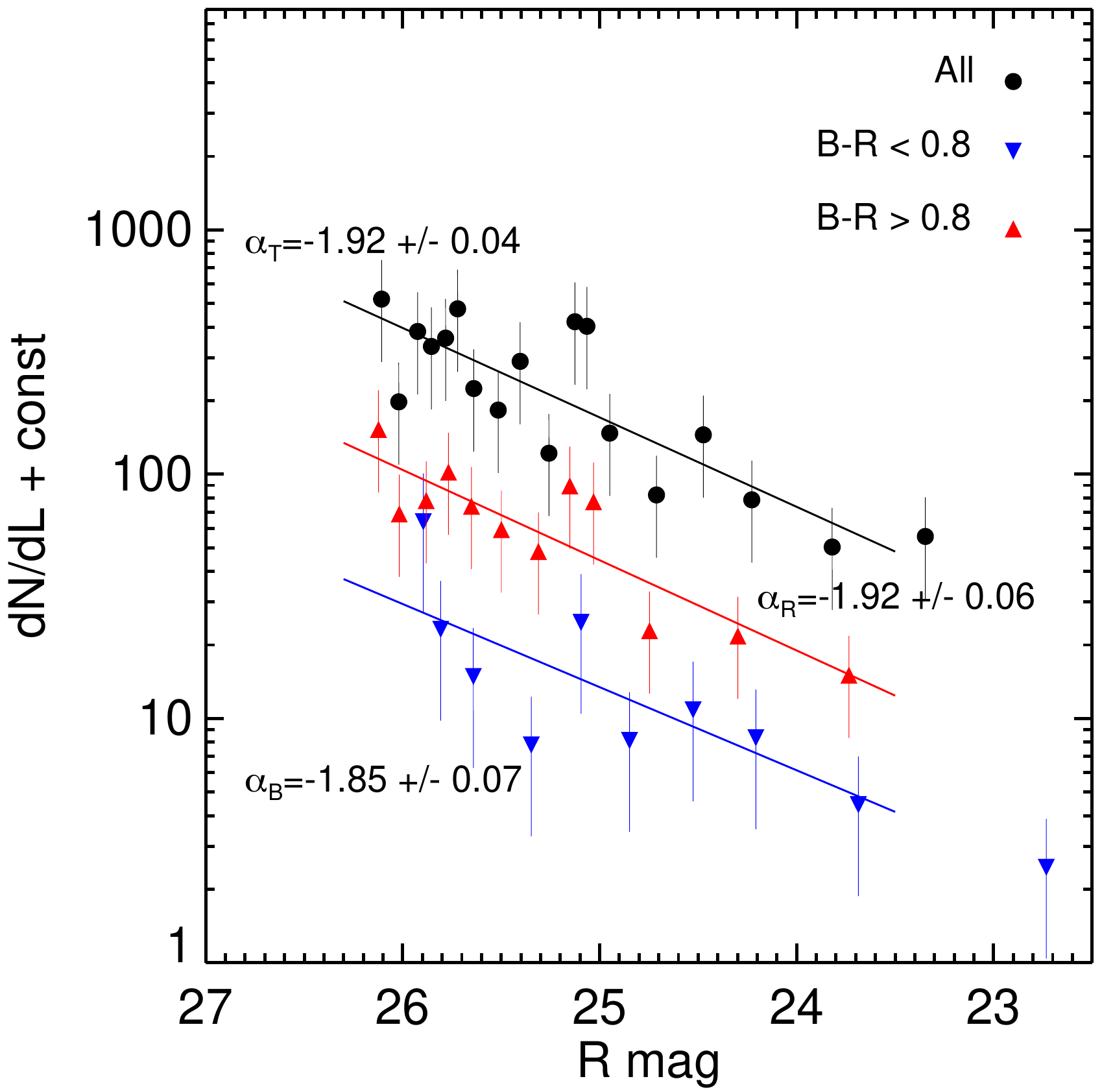}
	\caption{Luminosity function for all clusters detected in S12 (black circles), those bluer and redder than than $B-R$ of 0.8 (blue and red triangles, respectively). The best fit power-law index $\alpha$ is indicated for each distribution, and all of them are quite similar, with $\alpha\approx-1.9$. }\label{fig:dndl}
\end{figure*}

Young cluster populations in nearby galaxies are known to have luminosity (and mass) functions that can be represented by a power law, $dN/dL \propto L^{\alpha}$, with $\alpha\approx-2$ \citep[e.g.,][]{Harris&Pudritz94}. In Figure~\ref{fig:dndl}, we present the $R$-band luminosity function for the full cluster sample (black filled circles), as well as for blue $B-R < 0.8$ (blue triangles) and red $B-R > 0.8$ (red triangles) sub-samples, where $B-R=0.8$ separates clusters formed in the most recent burst from older ones. The plots show distributions with variable size bins and the same number of clusters in each bin, and the best fit, $dN/dL = \alpha \times~\mbox{mag} + \mbox{const}$, is plotted in each case. The luminosity functions for the entire cluster population, plus the blue and red sub-populations all have $\alpha\approx-1.9$, similar to results for young cluster populations found in normal galaxies, including irregulars, spirals, and mergers (e.g., \citealt{Fall12}).  

\subsection{Color-Color Diagrams} \label{subsec:2col}
A color-color diagram provides a useful guide to the ages of clusters, and insights into the amount of reddening, both foreground and internal to the system itself. There are a number of different stellar population models that predict the color evolution of stellar clusters. While most make very similar predictions for clusters older than $\gea 1$~Gyr, they show significantly more variation at younger ages, particularly $\tau \lea 50$~Myr. We have previously tested several different models, and found that overall, the \citet{Bruzual03}, hereafter BC03, population synthesis models best match the observed colors of young star clusters \citep[e.g.][]{Chandar10,Turner21}. In this work we adopt the BC03 stellar evolutionary models with solar metallicity, which is a reasonably good match to estimates for S12 \citep{French18}, but also check the results if we adopt the $1/4\times$solar model instead. 

\begin{figure*}[t]
    \centering
	\includegraphics[width=6.0in]{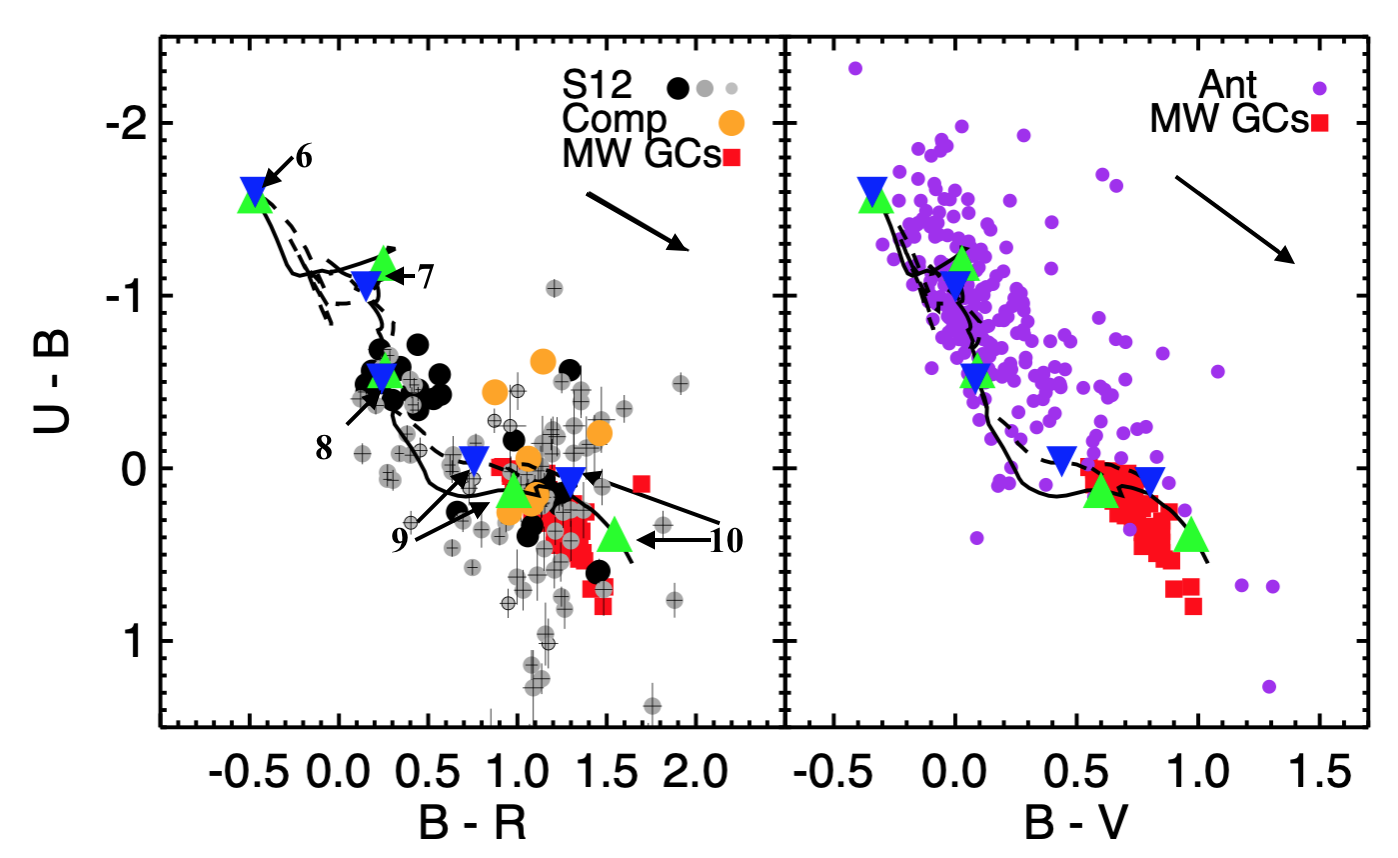}\label{2col}
	\caption{{\bf Left:} $B-R$ vs. $U-B$ color-color diagram of clusters in S12. The circle sizes indicate source luminosity ($<25.5$, $25.5-27$, $27-28$~mag), with larger circles representing brighter clusters. The rms uncertainty for each data point in S12 is shown in the left panel. Clusters detected in the companion are shown as orange circles and Galactic globular clusters as red squares. Both of these overlap with the colors of red clusters in S12. The solid line shows the predicted evolution of clusters from the Bruzual \& Charlot (2003) models with solar metallicity, starting from $10^6$ years in the upper left, down to $10^{10}$~yr. The $1/4\times$solar metallicity model is shown as the dashed line. Fiducial log~ages are indicated and marked for solar (green triangles) and $1/4\times$solar (inverted blue circles) metallicities, and the direction of reddening is indicated by the arrow. {\bf Right:} $B-V$ vs. $U-B$ two-color diagram showing the same models and Galactic globular clusters as in the left panel, but now including the measured colors for clusters (with $m_V\leq 21.5$~mag or $M_V\lea -9$~mag) in the merging Antennae galaxies (purple circles). The Antennae has clearly formed clusters more recently and continuously than S12. }\label{fig:colcol}
\end{figure*}

In the left panel of Figure~\ref{fig:colcol} we compare the measured $B-R$ vs. $U-B$ colors measured for clusters in S12 (filled circles), and its companion S12b (filled triangles), with predictions from the BC03 stellar population models (solar metallicity shown as the solid curve and $1/4$-solar as the dashed curve).
We see two `clumps' of clusters, a `blue' and `red' one; a few objects in each category are identified in Figure~\ref{fig:S12}. Brighter clusters in S12 are represented with larger symbols. The model ages start at 1~Myr in the upper left and end at 10~Gyr in the lower right, with every factor of 10 marked along the cluster evolution track. The solar and $1/4\times$solar metallicity models are fairly similar for ages up to $\sim0.5$~Gyr, but begin to diverge after this, when the well-known age-metallicity degeneracy begins to affect the broad-band colors of clusters.

Broad-band cluster colors can also be affected by reddening due to dust, both foreground and within the galaxy itself. The arrow shows the direction a cluster would move in this color-color space due to reddening, assuming a Milky Way-type extinction law (Fitzpatrick 1999). It is important to note that the measured colors for almost all of the clusters, particularly the brightest but even the fainter ones, fall right on top of the model predictions starting near $\approx100$~Myr and continuing to older ages, especially for the bluest clusters which are found near the center of the post-starburst galaxy. {\em This indicates there is little reddening towards clusters in S12}, which is not uncommon for clusters older than $\gea 10$~Myr \citep[see e.g.][]{Whitmore20}. It also suggests that the clusters are sufficiently massive that their colors are not affected by the presence (or absence) of just a few massive stars, as is typical for lower mass clusters \citep[ e.g.][]{Fouesneau10}. It appears that photometric uncertainties are responsible for the measured colors of the (few) clusters that lie somewhat off the model tracks.

In the left panel of Figure~\ref{fig:colcol}, we also show the dereddened $B-R$ vs. $U-B$ colors of globular clusters in the Milky Way from the catalog of \cite{Harris96} as the red squares\footnote{See \href{http://physwww.mcmaster.ca/$\sim$harris/mwgc.dat}{http://physwww.mcmaster.ca/$\sim$harris/mwgc.dat}}. These have very similar colors as the S12 clusters in the red `locus', and to clusters in S12b. In the right panel of Figure~\ref{fig:colcol} we compare the exact same BC03 model tracks but now for $B-V$ vs. $U-B$ so that we can compare with the cluster population in an on-going merger (many nearby star-forming galaxies have $V$ rather than $R$ band imaging). In this figure, in addition to Milky Way globular clusters, we also show the colors measured for clusters brighter than $M_V$ of $-9$ in the Antennae galaxies, a pair of spirals in the early stages of merging, that has a current star formation rate of $\approx20~M_{\odot}~\mbox{yr}^{-1}$ \citep{Whitmore20}. The on-going star and cluster formation in the Antennae is evident from the strong concentration of clusters near the youngest $\sim1$~Myr model colors, and quite different from the cluster population of S12. This more-or-less even distribution of cluster colors all along the model tracks is similar to that observed in other on-going mergers like NGC~3256 \citep[e.g.][]{Mulia16}, spirals like M51 and M83 \citep{Chandar14,Chandar17}, and even irregular and dwarf starburst galaxies like the Magellanic Clouds and NGC~4449 \citep[e.g.][]{Hunter03,Chandar10,Rangelov11}. We will return to the temporal relationship between S12 and other mergers in the Discussion section.
  
\subsection{Age \& Mass Estimates of Clusters}\label{subsec:Mt}
The ages of young stellar clusters can be estimated by comparing the measured luminosity in each filter with model predictions. In general this requires at least 3 filters, including the $U$ band, and an assumed metallicity. The details of how to implement this fitting typically depend on the number of available measurements, the range of reddening values expected in the galaxy, and the range of cluster masses (i.e. whether or not stochasticity is likely to have a strong impact). 

For the reasons mentioned in the previous section, the broadband colors measured for clusters in S12 do not appear to be affected by attenuation. Therefore to estimate the age of each cluster, we find the BC03 model which most closely matches it's measured $B-R$ and $U-B$ colors, i.e. we minimize the distance to the models in color-color space. We perform this procedure assuming both the solar and $1/4$-solar metallicity BC03 models. We find that if we had applied a correction for the foreground extinction prior to age dating, 80\% of the clusters have estimated ages which are essentially unchanged, and in only $4/115$ ($\sim3.5$\%) do the estimated ages change substantially, by $\gea50$\%. We find that these changes have a negligible impact on the mass functions and age distributions. 

The largest source of uncertainty in the age estimates likely comes from uncertainties in the photometry, particularly the $U$ band measurement for fainter clusters. Our estimates suggest that these uncertainties are fairly small for most clusters, and tend to be largest for the clusters which fall furthest from the models, as can also be seen in Figure~\ref{fig:colcol}.

We estimate the mass for each cluster as follows. The total $V$ band magnitude is estimated from the measured $R$-band magnitude and the predicted $V-R$ color at the best fit model age. We then combine the age-dependent mass-to-light ratio in the $V$ band ($M/L_V$) predicted by the BC03 model with the absolute V band magnitude of each cluster to estimate it's mass, assuming a \citet{Chabrier03} initial mass function.  The systematic errors are dominated by uncertainties in the assumed aperture correction (see Section~\ref{sec:obs}), distance to S12, and the assumed IMF. In general, mass estimates determined by comparing broad band colors with predictions from population synthesis models have been found to be uncertain by log~M$\approx0.3$ or a factor of $\sim2$ \citep{deGrijs05,Chandar10}. 

\section{Results}\label{sec:Results}

\subsection{Mass-Age Diagram}\label{subsec:massage}
\begin{figure*}[t]
	\centering
	\includegraphics[width=12.5cm]{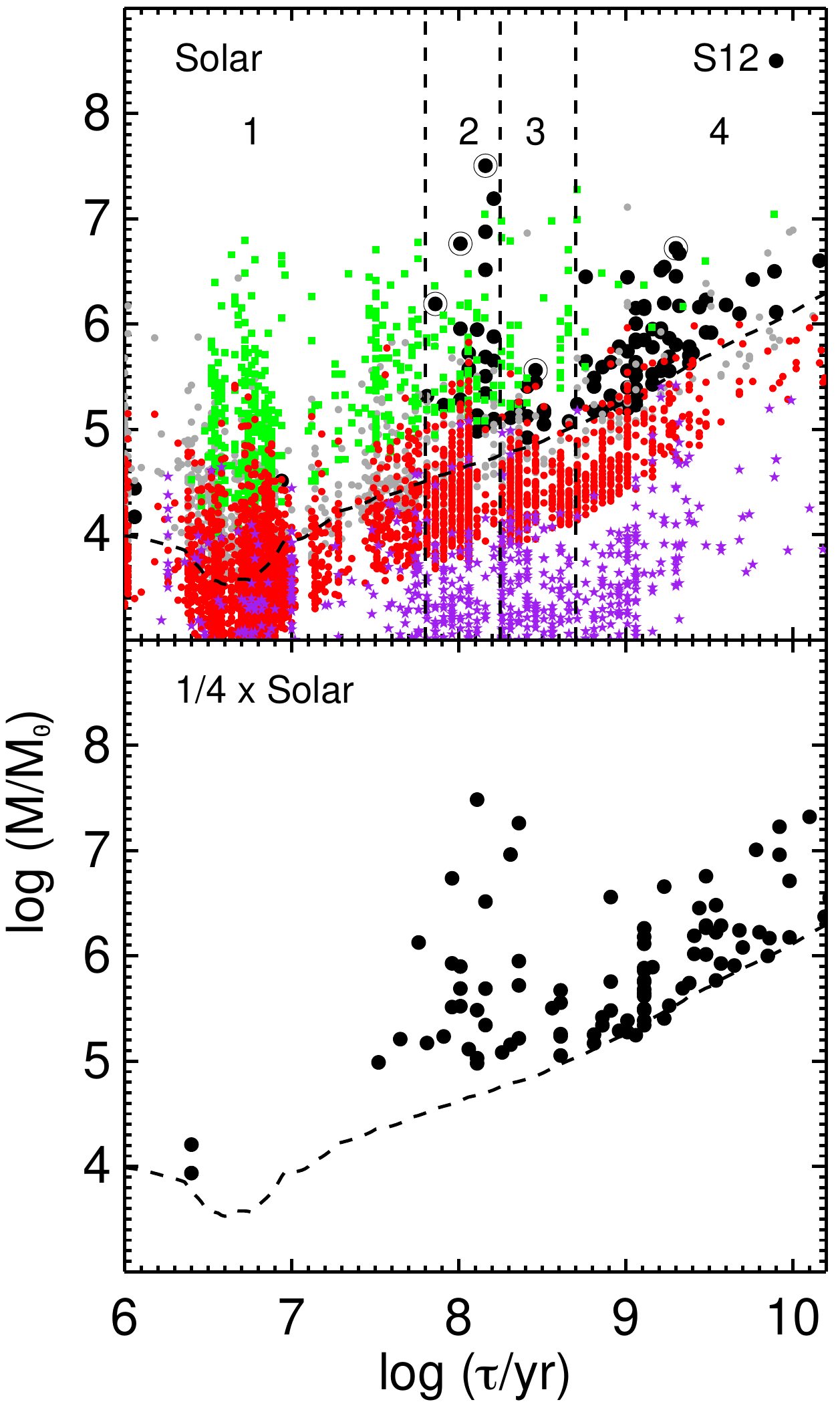}
	\caption{Mass-age diagram determined for clusters in S12 (solid black circles) when solar (top panel) and $1/4\times$solar (bottom panel) metallicity are assumed during the age dating process. The dashed line indicates how the completeness limit translates into the age-mass plane, and effectively shows the fading of clusters over time. We identify 4 distinct age intervals of interest based on the cluster demographics in S12, and circle the most massive cluster in each one. The top panel also shows the ages and masses determined for cluster populations in nearby galaxies with a range of star formation rates: NGC~3256 (green), Antennae (gray), M51 (red), and LMC (purple).}\label{fig:massage}
\end{figure*}

In Figure~\ref{fig:massage} we plot our age and mass results for the clusters, with assumed solar metallicity (top panel) and $1/4\times$solar metallicity (bottom panel). Solar metallicity is likely a better representation of the young cluster population. The solid curve indicates an approximate completeness limit of $M_V\sim-9$~mag, and effectively represents the fading of clusters over time, as the most massive and brightest stars die off. As expected, the main difference between the results for the two assumed metallicities shows up at ages $\gea0.5$~Gyr, with clusters having older estimated ages when the lower metallicity model is assumed. The results for younger clusters are more similar between the two metallicities, with the most massive young clusters having very similar ages in both cases.

The age-mass diagram (based on solar metallicity) over the past $\sim0.5$~Gyr can be divided into four distinct age intervals, indicated by the dashed vertical lines and labeled in the top panel of Figure~\ref{fig:massage}:
\begin{enumerate}
    \item There are no detected clusters younger than log~$(\tau/\mbox{yr}) \lea 7.8$ 
    \item The most massive clusters formed in the short time interval of  log~$(\tau/\mbox{yr})\approx 7.8-8.25$ 
    \item Clusters also formed in the age interval log~$(\tau/\mbox{yr})\approx 8.25-8.7$, but with significantly lower masses
    \item An older cluster population formed in S12, with a lower age limit of log~$(\tau/\mbox{yr}) > 8.7$. From the color-magnitude diagram we believe the most massive of these clusters are significantly younger than $\sim12$~Gyr; we constrain the ages of these clusters in Section~\ref{sec:oldsfh}.
\end{enumerate}

One intriguing possibility is that the SFR in interval~2 varied even over the short timescale between log~$(\tau/\mbox{yr})=7.8-8.25$, as possibly hinted at by the three clusters circled in this age interval in upper panel of Figure~\ref{fig:massage}. Here, it appears that the most massive clusters formed log~$\tau \sim8.1$~ago, but that the mass of the most massive cluster dropped significantly after that.  This implies that the burst of star formation had a very short duration, and decreased substantially before shutting off entirely.  We will return to this point below. 

Well-understood cluster populations in nearby star-forming galaxies provide an important reference for those in S12. In the top panel of Figure~\ref{fig:massage}, we include the mass-age results for cluster populations in two on-going mergers, NGC~3256 (green circles) and the Antennae (small gray circles), the spiral galaxy M51 (red), and the LMC, an irregular galaxy (purple). Unlike S12, {\em all} of these systems have had more-or-less continuous star and cluster formation over the past billion years, including up to the present day, since the data points continue almost all the way to the left edge of the diagram. The mass range of the cluster populations, however, is quite different between the different galaxies. The lower edge for each galaxy represents the luminosity-limit of each survey, which is mostly dictated by it's distance. Each cluster system should be fairly complete above this limit. At the upper end, we find that NGC~3256, the system with the highest rate of star formation, also has formed the most massive clusters at all ages, while the LMC, the system with the lowest rate of star formation (but similar mass to S12), has formed the least massive clusters. This correlation between star formation rate and cluster mass will be developed further in the next section, and used to estimate the star formation rate in the post-starburst galaxy S12 in the different age intervals identified here. 

\subsection{Star Formation Rate over the Last $0.5~Gyr$}\label{sec:recentsfh}
In this section, we estimate the star formation rate of S12 in age intervals 1, 2, and 3 from a new, empirical calibration between SFR and the most massive cluster $M_{\rm max}$ based on cluster populations in nearby, star-forming galaxies. The mass function of young cluster populations in nearby galaxies ($\tau \lea 0.5$~Gyr) has a near-universal shape, which is well-described by a power law, $dN/dM \propto M^{\beta}$, with $\beta=-2.0\pm0.2$ \citep[e.g.][]{Zhang99,Larsen02,Fall12}.\footnote{Although there have been claims that cluster mass functions in some galaxies may deviate downward (have an exponential cutoff) at the upper end, detection of an upper cutoff in the mass function does not appear to be statistically significant in most cases \citep{Mok19}.} In addition, the formation rate of the clusters appears to be proportional to the SFR, at least up to $\sim0.5$~Gyr \citep{Chandar17}. A direct consequence of this proportionality and universal shape is that the total number of clusters, and the mass of the most massive cluster, scales directly with the SFR of the host galaxy \citep[e.g.][]{Whitmore14,Chandar15}.

Because of the large distance to S12, the number of detected clusters in each age interval is fairly low, making it difficult to use the mass function directly (although we found that the luminosity function using all detected clusters is quite similar to those observed in nearby galaxies). Therefore, we develop a new method to estimate SFR based on the most massive cluster (and 3rd, 5th most massive clusters), building on the high quality data available for cluster systems in nearby galaxies, which cover a factor of $\sim5000$ in SFR.

\begin{figure*}[t]
	\centering
	\includegraphics[width=12.5cm]{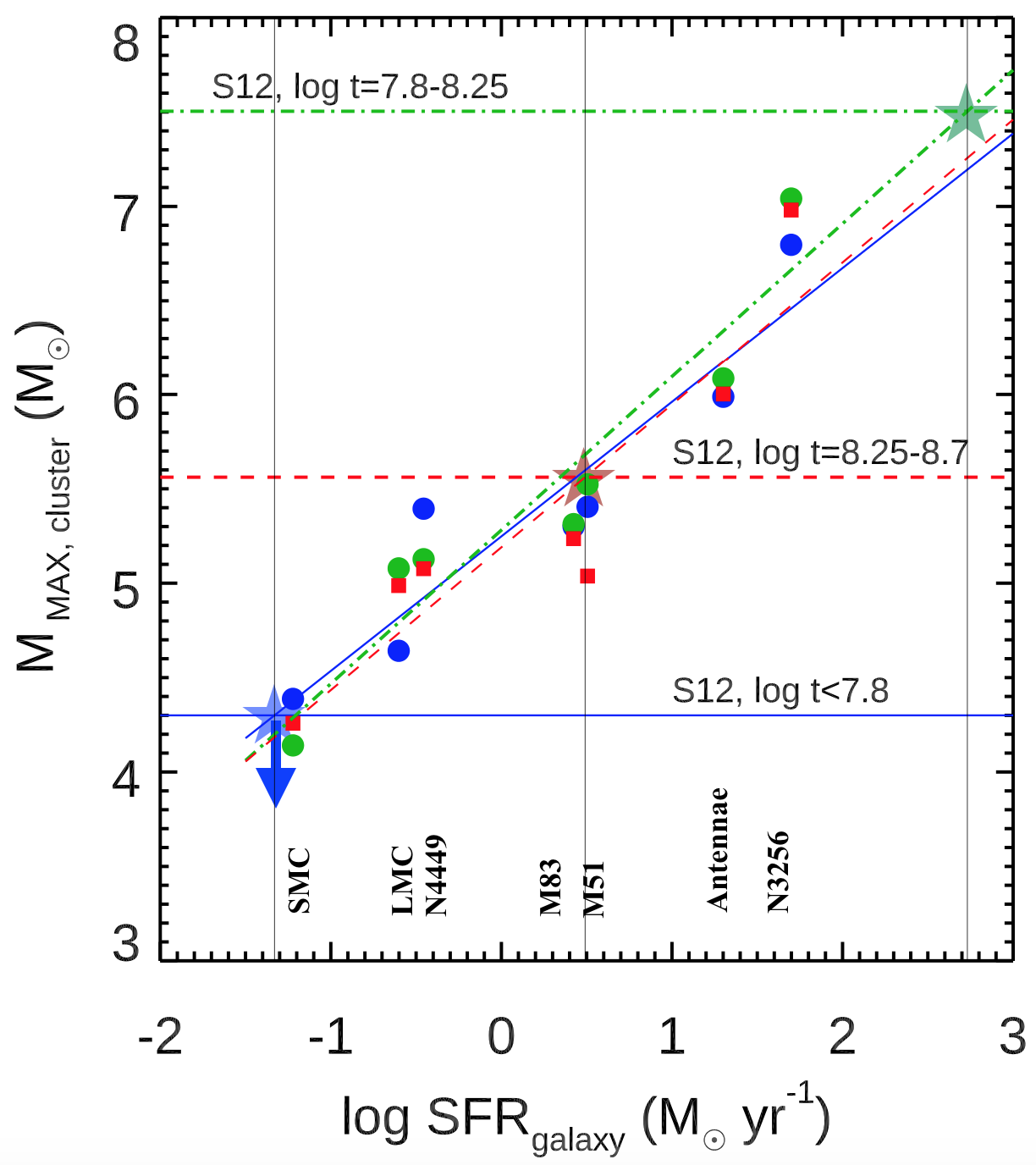}
	\caption{This figure shows the calibration between the SFR of the host galaxy and the most massive cluster ($M_{\rm max}$), based on 7 galaxies (LMC, SMC, NGC 4449, M83, M51, Antennae, and NGC~3256) with well-studied cluster populations. The calibration is done separately for each of three age intervals that are important in S12: (1) log~$(\tau /{\mbox yr})<7.8$ (blue circles), (2) log~$(\tau /{\mbox yr})=7.8 - 8.25$ (green circles), and (3) log~$(\tau /{\mbox yr})=8.25 - 8.7$ (red squares). The color-coded lines show the best fit relation for the nearby galaxies determined in Section~\ref{subsec:Mt} and \ref{sec:recentsfh}, and the most massive cluster found in S12 in each age interval is indicated by the horizontal lines. Stars show where these masses intersect the best fit calibration and the inferred star formation rate for S12. See text for details.} \label{fig:Mmax}
\end{figure*}

We illustrate our method in Figure~\ref{fig:Mmax}. We plot the SFR determined from extinction-corrected far ultraviolet luminosity (see Table~2 in \citealt{Chandar17}) against the mass of the most massive cluster in the LMC, SMC, NGC~4214, NGC~4449, M83, M51, the Antennae, and NGC~3256. This correlation is plotted separately for the three age intervals of interest for post-starburst galaxy S12: (1) log~$(\tau/\mbox{yr}) < 7.8$ (blue), (2) log~$(\tau/\mbox{yr}) = 7.8-8.25$ (green), and (3) log~$(\tau/\mbox{yr}) = 8.25-8.7$ (red). Figure~\ref{fig:Mmax} shows a strong linear correlation between log SFR and log~M$_{\rm max}$ for each age interval. We find the best fits to be:
\onecolumngrid
\vspace{10pt}
{\centering 
\noindent
$\mbox{log~M}_{\rm max} = (0.72\pm0.12) \times \mbox{log~SFR} + (5.2\pm0.1)~~~~~~~~~~~~~~~(\mbox{log}~(\tau/\mbox{yr}) = 6.0-7.8)$ \\
~~$\mbox{log~M}_{\rm max} = (0.81\pm0.10) \times \mbox{log~SFR} + (5.3\pm0.1)~~~~~~~~~~~~~~~(\mbox{log}~(\tau/\mbox{yr}) = 7.8-8.25)$ \\
~~$\mbox{log~M}_{\rm max} = (0.76\pm0.15 \times \mbox{log~SFR} + (5.2\pm0.1)~~~~~~~~~~~~~~~~(\mbox{log}~(\tau/\mbox{yr}) = 8.25-8.7)$ \\ }
\vspace{10pt}
\twocolumngrid
\noindent shown as the blue, green, and red lines, respectively. We estimate the SFR in S12 in each age interval by using the best fit relation given above and the estimated mass of the most massive cluster; these values are represented by the blue, green, and red horizontal lines. The estimated SFR can be visually estimated where the mass of the S12 cluster (horizontal line) intersects the best fit line of the same color. We repeat this entire procedure, starting with the calibration between log~SFR and log~M$_{\rm max}$ using the 3rd and 5th most massive clusters in each age interval (not shown), and find that the slopes are always within the fit uncertainties given above (as expected, the intercepts are lower, and nearly identical between the three age intervals). Below, we summarize our SFR estimates in each of the three age intervals.

\begin{itemize}
    \item {\bf log~$\tau< 7.8$}: no clusters were detected, so we can only place an upper limit on the SFR. Based on the luminosities of detected clusters, we conservatively estimate that we can detect clusters down to $M_R\sim-9$, which corresponds to a maximum cluster mass of log~$(M/M_{\odot}) \approx4.3$ over this age interval. Based on where the horizontal dashed blue line intersects the best fit calibration between the SFR and maximum cluster mass in Figure~\ref{fig:Mmax}, we estimate an upper limit for the SFR of $\lea 0.05~M_{\odot}~\mbox{yr}^{-1}$. This limit is consistent with the independent upper limit on the SFR of $\lea0.03~M_{\odot}~\mbox{yr}^{-1}$, based on H$\alpha$ measurements \citep{French18}.
    \item {\bf log~$(\tau/\mbox{yr})=7.8-8.25$}: The most massive cluster in S12 formed during this period has $M\approx 3\times10^7~M_{\odot}$, approximately a factor of 3 times more massive than its counterpart in NGC~3256. The calibration in Figure~\ref{fig:Mmax} gives an estimate of $\mbox{SFR}\sim500~M_{\odot}~\mbox{yr}^{-1}$, and the $1\sigma$ uncertainties on the best fit slope result in lower and upper limits of $\sim 250$ and $1000~M_{\odot}~\mbox{yr}^{-1}$. Estimates of the SFR from the 3rd and 5th most massive clusters also fall within this range. While the uncertainties are significant, the fact that there at least two clusters more massive than any found in NGC~3256 over this time period indicates that the SFR must have been quite high. We estimate the SFR of S12 during the time of its recent burst to be approximately $500^{+500}_{-250}~M_{\odot}~\mbox{yr}^{-1}$. The three most massive clusters in this interval all have nearly identical ages, log~($\tau/\mbox{yr})\approx8.1-8.25$. This may indicate that the most intense portion of the burst was extremely short lived, with a duration of only $\sim50$~Myr. Based on photometric and model uncertainties, we estimate errors in the burst duration of the order log~($\tau/\mbox{yr})\sim0.075$, or $50\pm25$~Myr. If we take seriously the declining mass of the most massive clusters (circled in Figure~\ref{fig:massage}, we can further divide interval~2 into the age ranges: log~($\tau/\mbox{yr})=8.0$-8.1 and 7.8-8.0. If we apply the best fit log~$M_{\rm max}$ vs. log~SFR relation found above to the most massive clusters and assume an uncertainty of a factor of 2, we estimate SFRs of $65^{+65}_{-33}~M_{\odot}~\mbox{yr}^{-1}$ (for log~($\tau/\mbox{yr})=8.0$ -8.1) and $13^{+13}_{-6}~M_{\odot}~\mbox{yr}^{-1}$ (for log~($\tau/\mbox{yr})=7.8 -8.0$).
    \item {\bf log~$\tau=8.25-8.7$}: The most massive cluster in S12 formed over this period has $M\approx 4\times10^{5}~M_{\odot}$. From this, the calibration gives $\mbox{SFR}\approx3.1\pm0.5~M_{\odot}~\mbox{yr}^{-1}$. The results from the 3rd and 5th most massive clusters are somewhat higher, closer to $\mbox{SFR}\approx5~M_{\odot}~\mbox{yr}^{-1}$. We adopt a SFR of $3\pm2~M_{\odot}~\mbox{yr}^{-1}$ for this time period.
\end{itemize}

\subsection{Constraints on the Star Formation Rate $>0.5$~Gyr Ago}\label{sec:oldsfh}
The brightest red ($B-R>0.8$) clusters in S12 are significantly more luminous than any known population of ancient globular clusters, which suggests that they formed more recently. In this section we constrain the ages of these younger `red' ($B-R>0.8$) clusters and attempt to estimate the SFR at the epoch of their formation.  

The broad-band colors of clusters older than $\gea1$~Gyr ($B-R > 0.8$) suffer from the well-known age-metallicity degeneracy \citep[e.g.,][]{Worthey93}, which makes it difficult to directly determine their ages from colors alone. However as noted earlier, the most luminous red clusters in S12 are significantly brighter (by nearly 2 magnitudes) than the brightest ancient ($\approx12$~Gyr) globular clusters in the Milky Way or any known galaxy. Taken together, their luminosities and colors provide important clues to their age. Based on their colors, the red clusters in S12 formed at least $\approx1$~Gyr ago, and based on their luminosities, they formed more recently than 12~Gyr ago. In fact, the brightest of these red clusters have luminosities similar to intermediate-age ($1-5$~Gyr) clusters discovered in post-merger galaxies like NGC~1316, where these intermediate ages have been confirmed spectroscopically \citep{Goudfrooij01}.
 
\begin{figure*}[t]
	\centering
	\includegraphics[width=11.5cm]{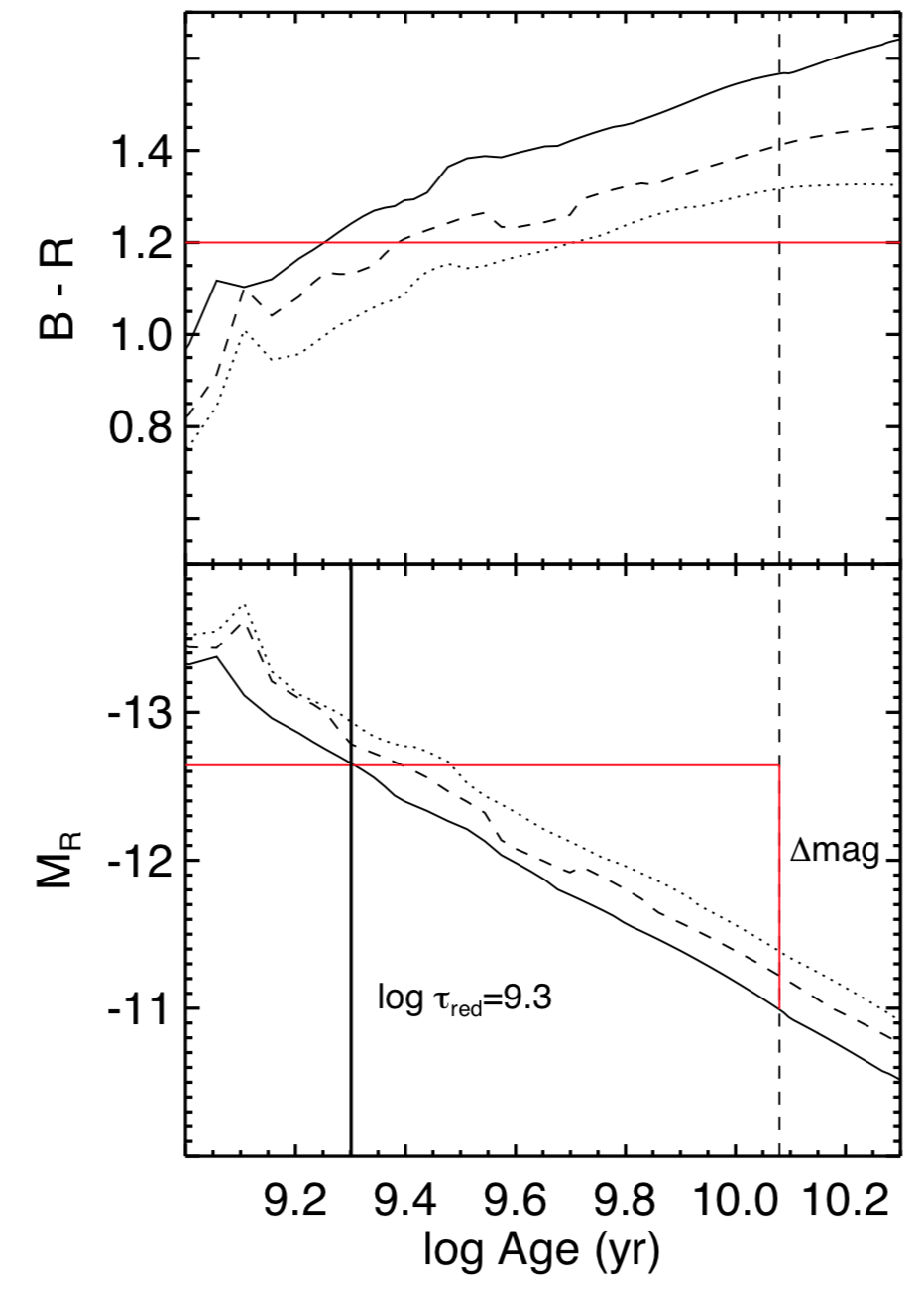}
	\caption{The predicted evolution of $B-R$ color (top panel) and $M_R$ brightness (bottom panel) at intermediate age clusters from the Bruzual Charlot (2003) models for solar (solid line), 40\% $\times$solar (dashed) and 20\% $\times$ solar (dotted) metallicity. The red horizontal lines show the median color of red clusters in S12 in the top panel, and the luminosity of the brightest red cluster in the lower panel. The normalization of the predicted luminosities in the lower panel are set by the difference in magnitude between $M_{\rm max}$ for the Galactic globular cluster and red S12 populations. This normalization to the observed properties of the Milky Way globular cluster system suggests that the red clusters in S12 have ages of log~$(\tau / {\rm yr})\approx 9.3$ or $\tau \sim 1.5$~Gyr. }\label{fig:GCevol}
\end{figure*}

Here, we follow a procedure similar to the one used by \citet{Goudfrooij01}, where the magnitude difference between the brightest red clusters in S12 and Milky Way globular clusters is used to infer the age of the former, based on predictions from SSP models. In Section~\ref{subsec:cmd}, we found this difference to be $\Delta_{\rm mag}=-1.7$ ($M_R=-12.5$ in S12 and $M_R=-10.8$ in the Milky Way). In the bottom panel of Figure~\ref{fig:GCevol}, we show the predicted $R$-band fading of a cluster between 1 and 14~Gyr, for 3 different metallicities: solar (solid line), $0.4\times$solar (dashed), and $0.2\times$solar (dotted), where we have normalized $M_R$ so that a 12~Gyr cluster (represented by the dashed vertical line) at solar metallicity has $M_R=-10.8$ to match the brightest globular cluster in the Milky Way from the Harris catalog. Clusters with the same mass but lower metallicity are brighter, as shown by the dashed and dotted lines. The red vertical line indicates $\Delta_{\rm mag}=-1.7$) starting at $M_R=-10.8$; we find that this $\Delta_{\rm mag}$ intersects the SSP model at an age of log~$\tau=9.3$ or 2~Gyr. This result does not depend on the assumed metallicity, since we would have found a similar result if we had normalized the $0.4\times$solar or $0.2\times$solar metallicity models to $M_R=-12.8$ instead.
  
There may be a difference in metallicity between the ancient globular clusters and intermediate age population. The top panel of Figure~\ref{fig:GCevol} shows that the median $B-R$ color of the intermediate age clusters in S12 falls between the blue and red globular cluster populations in the Milky Way. Taken at face value, this color indicates clusters with ages between $\sim1.7$~Gyr (solar metallicity) and $5$~Gyr ($0.2\times$solar metallicity). Based on the luminosity and color analysis presented above, {\em we believe that the bright red clusters in S12 constitute an intermediate age population, with an estimated age of $\tau \approx 2\pm1$~Gyr.}

Next, we attempt to constrain the rate of star formation $\sim2\pm1$~Gyr ago. While we know significantly less about intermediate age populations in nearby galaxies than those formed in the last $0.5$~Gyr, clusters in this age range are reasonably well studied in the LMC. We can estimate the SFR in S12 during this epoch by assuming that the intermediate age ($1-3$~Gyr) clusters follow a similar relation between M$_{\rm max}$ and SFR as their younger counterparts. \citet{Harris09} estimate the SFR in the LMC to be $\sim0.3-0.35~M_{\odot}~\mbox{yr}^{-1}$ by determining the effective temperatures of millions of stars. NGC~1783 is the brightest $1-3$~Gyr intermediate age cluster in the LMC, with $M_V=-8.2$~mag \citep{Bica96, Goudfrooij11}, $V-R\approx0.4$ (predicted from the $0.4\times$solar BC03 model), and hence $M_R\approx-8.6$~mag. These set the normalization for the scaling relation between SFR and M$_{\rm max}$ at this epoch. The difference in the R band magnitudes between the brightest intermediate-age cluster in S12 and NGC~1783 is $\Delta_{\rm mag}\approx-3.9$. For slopes between 0.72 and 0.82, which we found in the previous section, the calibration gives a peak SFR of $\approx 30-40~M_{\odot}~\mbox{yr}^{-1}$ during this epoch. 

\section{Discussion} \label{sec:discussion}

\subsection{Star Formation History of S12}\label{subsec:formation}

\begin{figure*}[t]
	\centering
	\includegraphics[width=12.5cm]{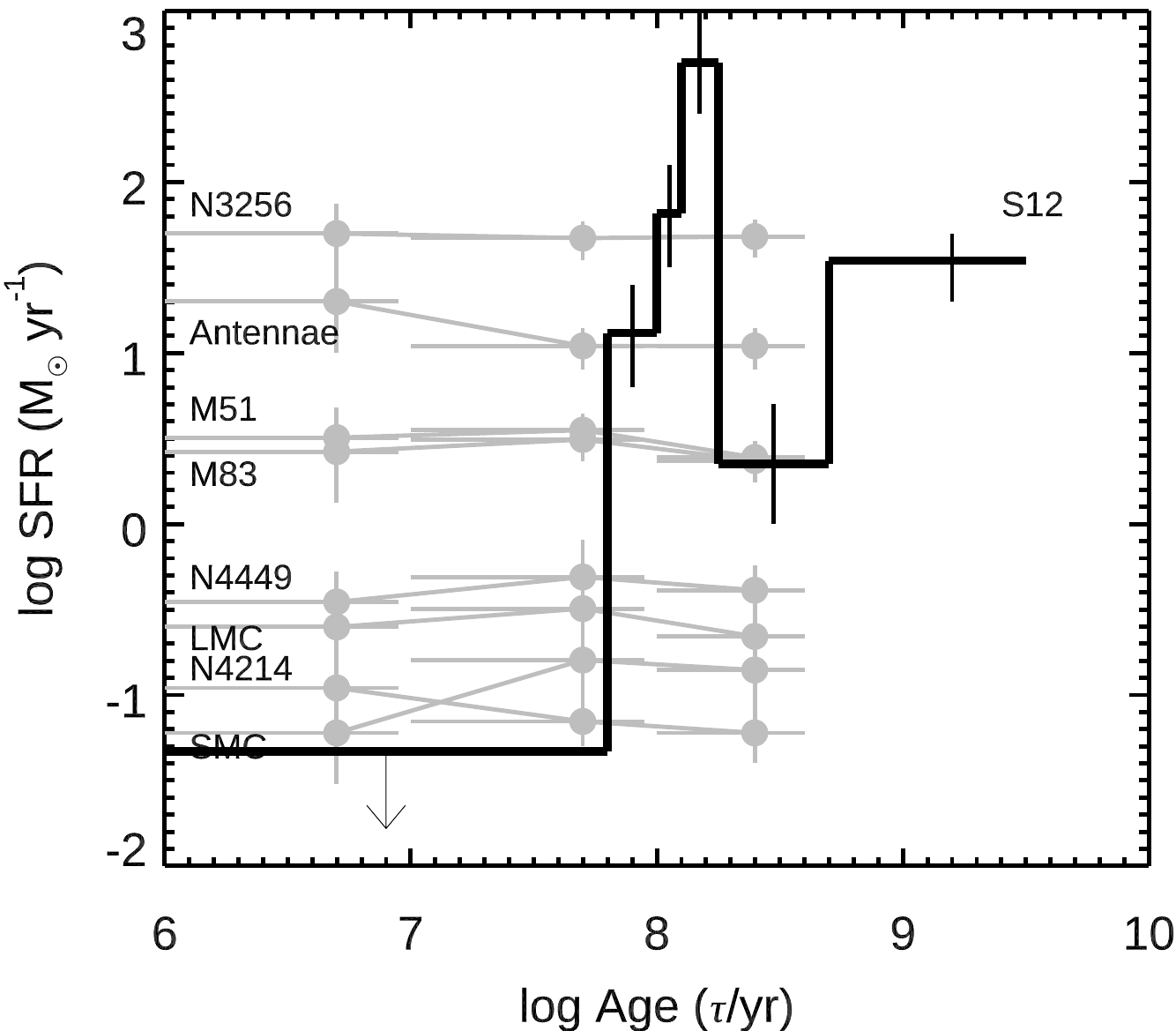}
	\caption{The recent star formation history of S12 determined from its cluster population, is shown as the black histogram. Those determined for the eight labelled galaxies from integrated light measurements (not clusters) and presented in \citet{Chandar17} are shown for comparison. The most recent star formation rate, from the present to $\sim70$~Myr ago is an upper limit, since no clusters which formed more recently have been detected. The peak SFR estimated for S12 is significantly higher than found for on-going mergers like the Antennae and NGC~3256.}\label{fig:sfh}
\end{figure*}

The star formation history of S12 provides clues to the processes that drove the system into it's post-starburst state. Some key constraints are the number, intensity, and duration of recent bursts of star formation. In Figure~\ref{fig:sfh}, we plot the star formation history of S12 (in logarithmic bins) based on our analysis of the cluster population in Section~\ref{sec:Results}; in Figure~\ref{fig:sfhcomp} we show the recent star formation history of S12 in linear bins. Our cluster analysis indicates that there has been little to no detectable star/cluster formation in the last $\approx70$~Myr, with a conservative upper limit of $\lea 0.05~M_{\odot}~\mbox{yr}^{-1}$, consistent with upper limits from other tracers (see Section~\ref{subsec:comp-intlight}). This cessation took place in the aftermath of an intense but short-lived burst, which reached an estimated peak star formation rate of $500^{+500}_{-250}~M_{\odot}~\mbox{yr}^{-1}$. The estimated duration of this intense period is only $\approx25$~Myr. To contextualize the intensity of S12's starburst, we compare it to the LMC. While best estimates put S12 at a comparable stellar mass to the LMC (${\sim}2{\times}10^9~M_{\odot}$), its peak SFR, though only briefly maintained, was $\sim$1000$\times$ higher than estimates of the LMC's peak SFR throughout its \textit{entire} interaction with the Milky Way (see Figure \ref{fig:sfh}). S12's implied specific SFR (sSFR) of ${\sim}2{\times}10^{-7}~{\rm yr}^{-1}$\ at that time is at least an order of magnitude higher than any galaxy in the low-redshift universe. This incredible system provides a truly unique window onto merger-driven galaxy evolution.

Interestingly, S12 experienced an earlier episode of enhanced star formation (i.e., above that expected for typical star forming galaxies) at some point between $\approx1$ and $\approx5$~Gyr ago, which we were only able to identify because the clusters formed during this episode are significantly brighter than any known ancient globular cluster. Cluster populations with these intermediate ages are generally difficult to identify and study because their colors overlap with those of the ancient clusters found in nearly all galaxies. This population of clusters in S12 has nearly identical luminosities and colors as the (spectroscopically confirmed) intermediate age population in the post-starburst galaxy NGC~1316. Although we cannot place any constraints on the {\em duration} of the star-forming episode, we were able to very roughly estimate the peak rate of star formation by extrapolating from the well-studied intermediate-age cluster population in the LMC. This extrapolation gives $30\pm15~M_{\odot}~\mbox{yr}^{-1}$, comparable to the current rate of star formation in the Antennae galaxies, which is in an early stage of merging having experienced it's first closest passage only a few hundred Myr ago \citep[e.g.][]{Whitmore99,Karl11,Chandar17}. After this episode sometime between $1-5$~Gyr ago, S12 settled down to form stars at a more moderate rate typical of spiral galaxies, a few solar masses per year, before experiencing the recent short, intense burst that drove the system into its post-starburst phase. The strong bursts of star formation and rapid changes in rate experienced by S12 over the past few billion years suggest that a major merger was responsible for driving this system into its post-starburst phase. 

Simulations and observations of nearby, gas-rich galaxy mergers suggest that
during the first close passage, when the original galaxies are still distinct, star and cluster formation is spatially distributed since it occurs in the native spiral disks, as observed in the Antennae, a system just past its first approach {\bf (e.g., \citealt{ Renaud14,Renaud15,Whitmore10})}. Eventually, violent relaxation will further distribute the clusters that are currently forming. As the merger progresses, star formation will become more spatially concentrated to the center of the system \citep[e.g.][]{Barnes04,DiMatteo07,Bournaud08, Saitoh09,Teyssier10}, which in some cases can lead to an extremely strong star forming event that expels the ISM and shuts off star formation altogether. \citet{Renaud15} presented the star formation history from a hydrodynamical simulation of an Antennae-like galaxy merger at parsec resolution, which included a multi-component model for stellar feedback. In that system, they found a final burst of centrally concentrated star formation with a duration of $\sim50$~Myr during the final coalescence of the merging galaxies, quite similar to the timescale found here for the most recent, intense burst. 

This general picture of star formation during a merger should lead to a signature in the locations of clusters with different ages, where the most recently formed clusters are more centrally concentrated than those formed earlier in the interaction. 
In Figure~\ref{fig:location} we show that the clusters in S12 follow this expected trend. Clusters formed in the recent burst $\sim70-170$~Myr ago are shown in blue, those formed $\approx170-500$~Myr are shown in green and those formed $\gea1$~Gyr ago are shown in red. The recently formed clusters are extremely concentrated towards the center of the system, whereas the older clusters are much more evenly distributed around the entire system. We conclude that the star formation history and the spatial distribution of the clusters in S12 support a merger origin for this post-starburst system. The morphology of the debris (e.g., fan/shell appearance) also suggests an earlier merger with a highly radial infall \citep{Johnston08}.

\subsection{Comparison with Composite Light Fitting Results}\label{subsec:comp-intlight}
\begin{figure*}[t]
	\centering
	\includegraphics[width=13.5cm]{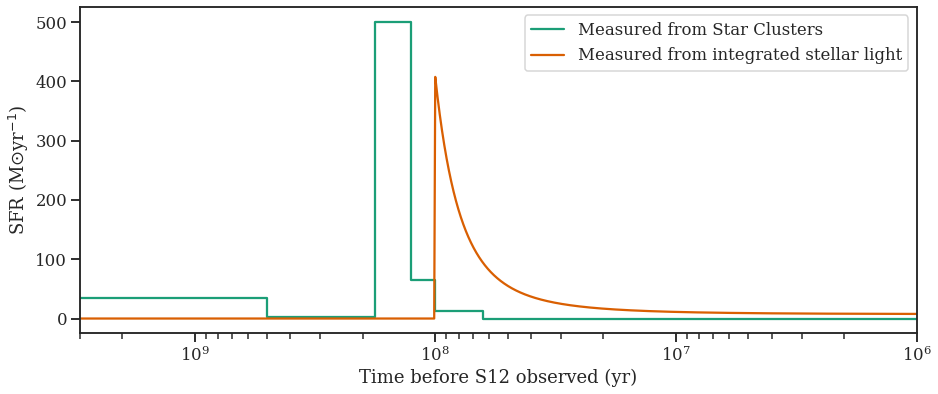}
	\caption{A comparison of the star formation history of S12 determined from stellar clusters (green) and composite light fitting (orange). The present day is located at $10^6$~yr, at the right end of the horizontal axis. Note the good agreement between the two independent methods over the most recent 0.5~Gyr, the time period for which the composite light fitting is sensitive.}\label{fig:sfhcomp}
\end{figure*}

Most of the information we have about post-starburst galaxies has come from fitting their composite starlight, including photometry and optical spectra. \citet{French18} modelled the integrated light of several hundred post-starburst systems, including S12. They assumed each system experienced an early star-forming event (which presumably formed globular clusters some $\sim10-12$~Gyr ago), plus either one or two bursts of star formation within the past Gyr (modeled by decaying exponentials). For systems which are best fit by a single burst, the age and burst duration are constrained; for systems best fit by two bursts, the separation in time between the two is constrained. For S12, \citet{French18} fit \textit{GALEX} FUV and NUV and SDSS \textit{ugriz} photometry, plus Lick indices measured from SDSS optical spectra of the central 3$\arcsec$, and found this system is best fit by a strong, single burst which had a duration of $\approx25$~Myr and began around $84-110$~Myr ago. No estimate of the uncertainty in the burst duration is given, but S12 has one of the shortest durations found in their sample. The composite star light fitting does not give any constraints or upper limits on the current SFR.

In Figure~\ref{fig:sfhcomp}, we compare the SFH results from the composite light fitting from \citet{French18} with that determined here from the clusters. This figure inverts the timeline from the previous ones to run from the past to the present. The derived properties for the recent burst, including approximate timing, intensity and shape, are remarkably similar from the two independent methods. The most obvious difference is the somewhat older peak found from age-dating the clusters compared with the composite spectral fitting. This offset is almost certainly due to the different assumptions made by the two works regarding extinction. In this work, we have assumed that extinction does not affect the clusters at all, since their photometry closely follows the model tracks. The composite star light fitting has extinction as a free parameter, and finds $A_V=1.2$~mag or $E_{\mbox{B-V}}=0.38$~mag for S12. We find that if we deredden the measured cluster colors by this same amount (assuming a Milky Way extinction law), the peak in star formation as measure by cluster ages and masses would be shifted closer to $\sim$80~Myr, nearly identical to the results from the composite fits.\footnote{While applying $E_{\mbox{B-V}}=0.38$~mag to the measured cluster colors brings their ages into better agreement with those determined from composite starlight fitting, it over-corrects the colors for most clusters, moving them off the model predictions, and leading to poorer quality fits.}

\begin{figure*}[t]
	\centering
	\includegraphics[width=13.5cm]{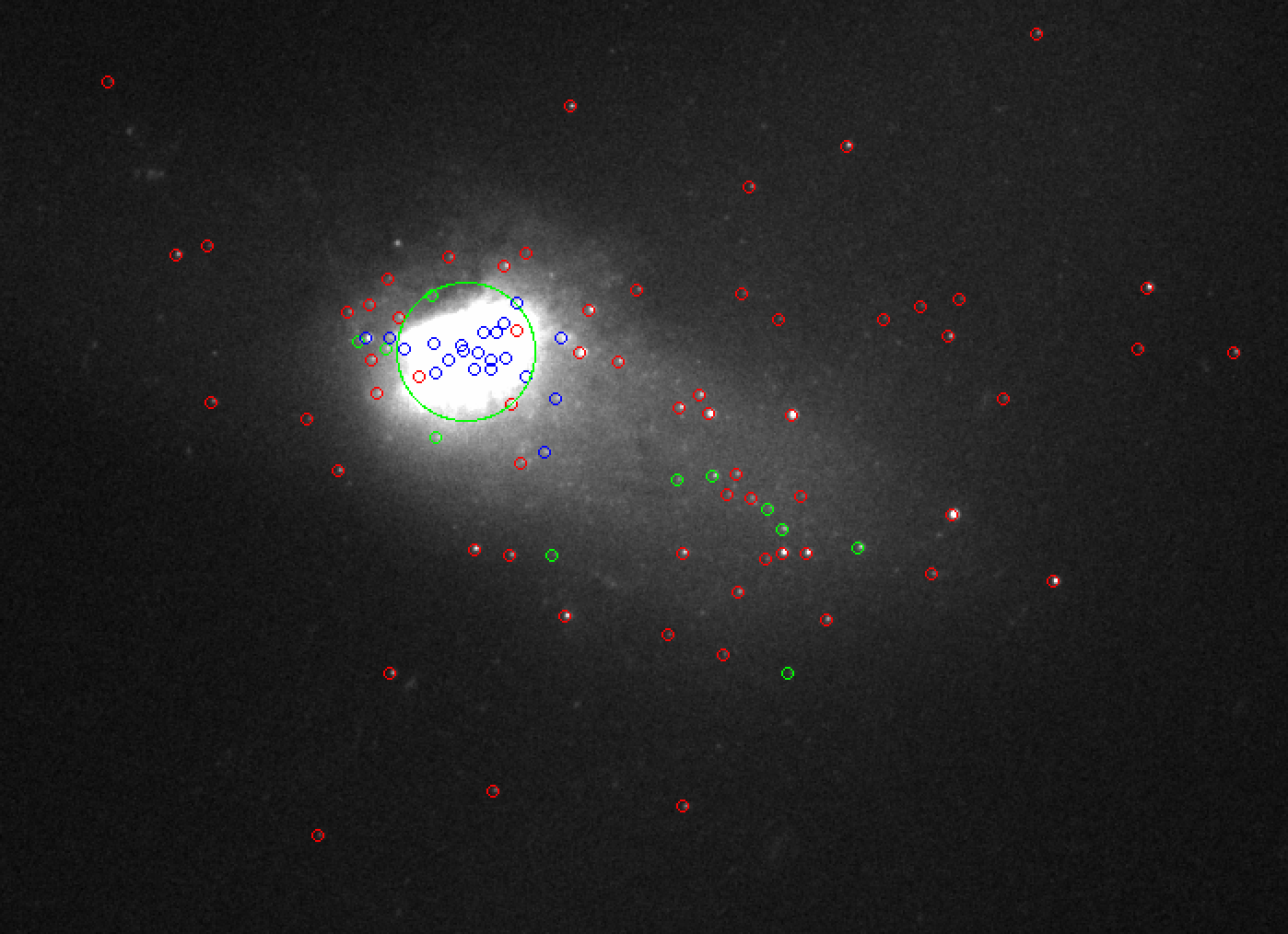}
	\caption{The locations of clusters in the 3 age intervals of interest in S12: log~$(\tau / {\mbox yr})=7.8-8.25$ (blue), log~$(\tau / {\mbox yr})=8.25-8.7$ (green), log~$(\tau / {\mbox yr})>8.7$ (blue). The youngest clusters are almost entirely concentrated in the bright central region, whereas the oldest clusters are much more spread out, with almost none found in the bright central portion of S12. The intermediate-age clusters have a somewhat intermediate-spatial distribution between their young and old counterparts. 
	The green circle shows that the SDSS fiber covers young clusters almost exclusively. }\label{fig:location}
\end{figure*}

The composite light fitting may be somewhat affected by the limited spatial coverage of the SDSS fiber, shown in Figure~\ref{fig:location} as the large green circle. Clusters which fall within the fiber nearly all formed during the recent burst, with older clusters found preferentially outside of this central region. The partial coverage and tendency for older clusters to reside further from the center of S12 may explain why the composite light fitting didn't detect the intermediate age stellar population represented by the red clusters, although the integrated light measurements are generally less sensitive to older stars. Most post-starbursts studied in the \citet{French18} sample do not suffer from limited spatial coverage because they are mostly more distant than S12.

There have been a number of independent composite light-based estimates made of the current rate of star formation in S12, based on different tracers and techniques. In this work, we used the fact that no clusters younger than $\lea 80$~Myr are detected down to our completeness limit to set an upper limit on the current SFR of $<0.05~M_{\odot}~\mbox{yr}^{-1}$. Estimates using other tracers are all consistent with this. The lack of [Ne\,\textsc{ii}]$+$[Ne\,\textsc{iii}] lines in the Spitzer spectrum gives an upper limit of $<0.14~M_{\odot}~\mbox{yr}^{-1}$, the [C\,\textsc{ii}] 158\textmu m line give $<0.3~M_{\odot}~\mbox{yr}^{-1}$, a measurement of the total infrared gives $0.35~M_{\odot}~\mbox{yr}^{-1}$ \citep{Smercina2018}, and 1.4~GHz measurements give $<1.07~M_{\odot}~\mbox{yr}^{-1}$. PAHs are a poor star formation rate indicator in this system, because of its unusual (bursty) star formation history. 

\subsection{Constraints from the ISM}\label{subsec:ism}

\begin{figure*}[t]
	\centering
	\includegraphics[width=0.82\linewidth]{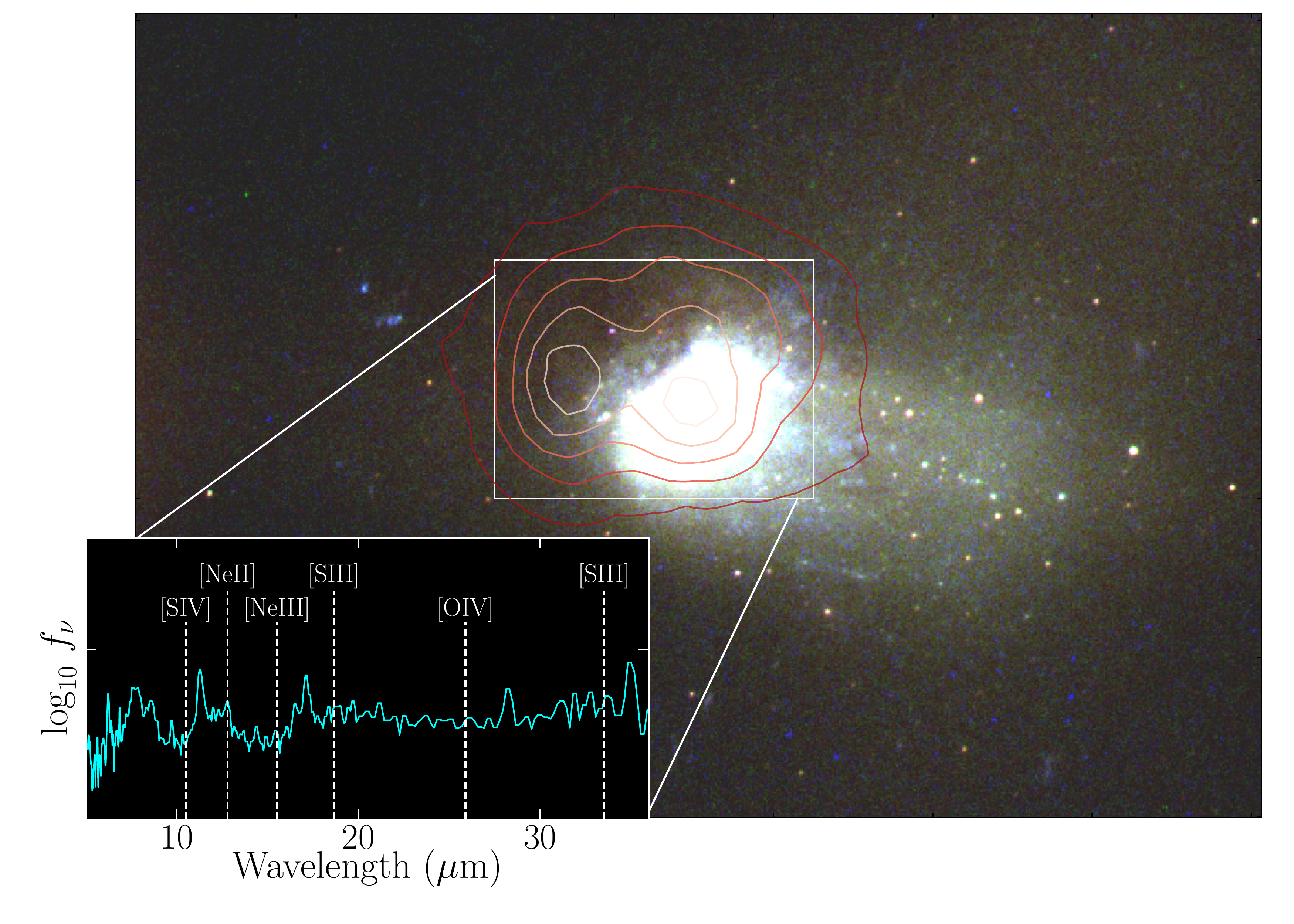}
	\caption{8\textmu m contours overlaid on an optical image of S12 reveals that roughly half of the strongest emission is off-nuclear, and originates from a portion of the galaxy that is optically dark. Extracted spectra from the \textit{Spitzer} IRS and covering the wavelength range 5--36\,\textmu m, show no visible nebular emission (lower-left inset; \citealt{Smercina2018}). This lack of mid-infrared nebular emission suggests that the $8\mu$m emission does not indicate highly obscured or embedded recent star formation. See Section~\ref{subsec:ism} for more details.}\label{fig:ism}
\end{figure*}

The intense burst of star formation that ended $\sim70$~Myr ago and shut off star formation in S12 did not expel all of the ISM. PAHs have been detected in observations taken by Spitzer, cold molecular gas traced by CO and bright pure rotational molecular hydrogen emission have been observed with IRAM. These observations give estimates of $\sim3\times10^9~M_{\odot}$ of kinematically hot molecular gas, $2\times10^7~M_{\odot}$ of dust, and $3.5\times10^8~M_\odot$ of cold molecular gas \citep{French2015,Smercina2018}. The contours in Figure~\ref{fig:ism} show two peaks of 8 \textmu m emission, the weaker one coincident with the brightest optical portion of S12. The strongest 8 \textmu m emission, meanwhile, comes from a location to the east of the optically bright central region of S12, where there is little optical emission and no detected clusters.

Is it possible that there is some highly obscured, very recent star formation hidden at the locations of the 8\textmu m peaks? We extracted Spitzer spectra (inset in Figure~\ref{fig:ism}), which cover the weaker, central peak of 8\textmu m emission and extend just to the stronger peak without covering it fully. Neither spectrum shows any NeII emission lines or Silicate absorption features, as we would expect if these sites were actively forming stars. Simulations have shown that infrared and PAH emission does not necessarily indicate active star formation in galaxies with unusual star-formation histories, particularly those where star formation has recently been quenched \citep[e.g.][]{Hayward14,Utomo14,Smercina2018}.

We conclude that despite the presence of molecular gas and dust, S12 is not currently forming stars, although it is possible this system may `come back to life' and begin a new round of star formation in the future. Since current star formation is not responsible for the 8~\textmu m emission, the gas is likely heated thermally and by the 70~Myr stellar population.

\section{Conclusion} \label{sec:conclusion}
To date, almost all determinations of the star formation histories of post-starburst galaxies have come from the  analysis of integrated starlight. While this approach can identify systems that have experienced a single burst versus those that have experienced a double burst in the last billion years, this method does not provide strong constraints on the rate or duration of the star forming event(s), or on the star formation history at intermediate ages, $\approx1-5$~Gyr ago.

In this work we have taken a different approach. For the first time, we have used the age and mass estimates of star clusters from multi-band $HST$ images of a post-starburst galaxy to determine its recent ($\lea 3$~Gyr) star formation history. The star formation rates in different age intervals are determined from a new calibration developed between the well-studied cluster masses and star formation rates in 8 nearby galaxies. The methods used here could be applied to determine the recent star formation histories for a number of post-starburst systems. 

The post-starburst S12 has a jellyfish-like structure, and a nearby companion (S12b). We detected and measured the brightnesses of 115 point-like clusters embedded in a background of diffuse light from the system. The luminosity function of these clusters can be described by a power law, $dN/dL \propto L^{\alpha}$, with $\alpha\approx-1.9$, similar to that found for cluster systems in nearby spiral, irregular, and merging galaxies. We compared the measured colors and luminosities of the clusters with predictions from the Bruzual \& Charlot stellar evolutionary models to estimate their ages and masses, and found that the age-mass diagram can be divided into four distinct age intervals: (1) log~$(\tau/\mbox{yr}) \lea 7.8$; (2) log~$(\tau/\mbox{yr}) = 7.8-8.25$; (3) log~$(\tau/\mbox{yr})=8.25-8.7$; and (4) log~$(\tau/\mbox{yr}) >8.7$. Our main results are:

\begin{enumerate}
    \item We found no clusters which formed in the last log~$(\tau/\mbox{yr}) \lea 7.8$. Based on the detection limit, we set an upper limit on the current rate of star formation of $<0.05~M_{\odot}~\mbox{yr}^{-1}$.
    \item Just before this cessation, S12 experienced one of the most intense bursts of star formation known in the nearby universe, forming stars at an estimated peak rate of $500^{+500}_{-250}~M_{\odot}~\mbox{yr}^{-1}$ for a very short period of time approximate 150~Myr ago. The cluster population also tentatively suggests that this rate of star formation decreased down to $65^{+65}_{-33}~M_{\odot}~\mbox{yr}^{-1}$ approximatly 100~Myr ago, and to $13^{+13}_{-6}~M_{\odot}~\mbox{yr}^{-1}$ approximately 70~Myr ago, after which it was abruptly truncated, with a current SFR of $\lesssim 0.1~M_{\odot}~\mbox{yr}^{-1}$.
    \item Leading up to the intense burst, the star formation rate of S12 was just a few solar masses, per year, typical of nearby spiral galaxies.
    \item S12 formed an earlier, now intermediate age cluster population. While such populations have been found in other post-mergers like NGC~1316, their ages and star formation rates are harder to constrain. We estimate these clusters formed during a star-forming event that took place sometime between $\approx1$ and 3~Gyr ago, and may have had a SFR that reaches as high as $\approx30-40~M_{\odot}~\mbox{yr}^{-1}$.
\end{enumerate}

We suggest that S12's unusual SFH resulted from a merger, which shut off star formation. The spatial distribution of clusters of different ages, where the youngest sources are much more centrally concentrated than the older ones, also supports a coalescing merger. This intense burst did not, however, expel all of the ISM. We examine Spitzer spectra at the peak 8\textmu m locations, and do not find any NeII or other emission lines indicating these are sites of deeply buried star formation. Rather, the ISM appears to be heated by the older stars that are present.

Our results for the timing of the recent burst based on an analysis of the cluster population is in excellent agreement with composite starlight fit results. The composite starlight fits did not detect the older burst which occurred at some point $\approx 2\pm1$~Gyr ago. \\

R.C.\ acknowledges support from NSF grant 1517819. A.S.\ is partially supported by NASA through grant \#GO-14610 from the Space Telescope Science Institute, which is operated by AURA, Inc., under NASA contract NAS 5-26555.

\software{\texttt{Photutils} \citep{photutils}, \texttt{Matplotlib} \citep{matplotlib}, \texttt{NumPy} \citep{numpy-guide,numpy}, \texttt{Astropy} \citep{astropy}, \texttt{SciPy} \citep{scipy_new}, \texttt{SAOImage DS9} \citep{ds9}}

\bibliographystyle{aasjournal}

\end{document}